\documentclass[aip,rsi,reprint]{revtex4-1}
\usepackage{graphicx}
\usepackage{amssymb}
\begin{document}

\title{The Intermodulation Lockin Analyzer} 

\author{Erik A. Thol{\'e}n}
\affiliation{Intermodulation Products, Stockholm, Sweden}
\affiliation{Nanostructure Physics, Royal Institute of Technology (KTH), Stockholm, Sweden}

\author{Daniel Platz}
\affiliation{Nanostructure Physics, Royal Institute of Technology (KTH), Stockholm, Sweden}

\author{Daniel Forchheimer}
\affiliation{Nanostructure Physics, Royal Institute of Technology (KTH), Stockholm, Sweden}

\author{Vivien Schuler}
\affiliation{Nanostructure Physics, Royal Institute of Technology (KTH), Stockholm, Sweden}
\affiliation{ENS Cachan, Paris, France}

\author{Mats O. Thol{\'e}n}
\affiliation{KUAB, R{\"a}ttvik, Sweden}

\author{Carsten Hutter}
\affiliation{Intermodulation Products, Stockholm, Sweden}
\affiliation{Nanostructure Physics, Royal Institute of Technology (KTH), Stockholm, Sweden}

\author{David B. Haviland}
\affiliation{Nanostructure Physics, Royal Institute of Technology (KTH), Stockholm, Sweden}
\email[]{haviland@kth.se}
\homepage[]{http://www.nanophys.kth.se}



\date{\today}

\begin{abstract}
Nonlinear systems can be probed by driving them with two or more pure tones while measuring the intermodulation products of the drive tones in the response.  We describe a digital lock-in analyzer which is designed explicitly for this purpose.  The analyzer is implemented on a field-programmable gate array, providing speed in analysis, real-time feedback and stability in operation.  The use of the analyzer is demonstrated for Intermodulation Atomic Force Microscopy.  A generalization of the intermodulation spectral technique to arbitrary drive waveforms is discussed.  
\end{abstract}

\pacs{}

\maketitle 

Nonlinearity is essential for the function of digital circuits and amplifiers, and yet it is often considered the bane of electronic systems, causing undesirable frequency mixing known as intermodulation distortion (IMD).  In some cases however, IMD can be useful for signal conditioning, for example in the phase-sensitive non-degenerate parametric amplifier, where a strong pump tone is used to deliver power gain to a weaker signal tone \cite{yurke:joseph-par-amp:89,Tholen:WeakLink:09}.  IMD can also be considered as useful if one is trying to gain understanding about the nonlinearity itself, for example in Intermodulation Atomic Force Microscopy \cite{Platz:ImAFM:08}.  Whether one wants to use or avoid nonlinearity, a complete understanding of the nonlinearity is often desired, and to this end we developed a digital lockin analyzer for probing the real-time intermodulation response of a nonlinear two-port system.    

The modeling and parametrization of nonlinearity is the object of current microwave instrumentation development \cite{Vye:Xparameters:10}, where there exists an engineering need for accurate models of the large amplitude behavior of microwave devices.  The vector network analyzer, the microwave version of the lockin, has recently been adapted for this purpose to measure the so-called X-parameters.  This measurement is based on the poly-harmonic distortion model \cite{Versprecht:polyharmonic:06}, which assumes the existence of a strong drive tone, and approximates the mixing of weaker signals which are harmonics of the drive tone with a linear mapping in frequency space.  The lockin analyzer discussed here is designed to measure intermodulation, or the frequency mixing of two or more drive tones applied to a nonlinear system.  

\begin{figure}
\includegraphics{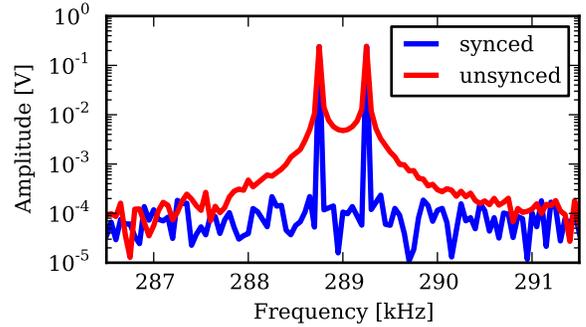}
\caption{The frequency spectrum of two pure tones is measured by sampling and discrete Fourier transform.  The large background near the peaks (red curve) arises from the removal of synchronization between the clock in the waveform synthesizer and the sampling clock.}
\label{sync}
\end{figure}

\section*{Lockin Measurement, Harmonics and Intermodulation}

The lockin measurement technique is defined by its use of a reference oscillation for making a narrow-band measurement of some physical process \cite{Dicke:Radiometer:46}.  The digital intermodulation lockin analyzer (ImLA) described here is capable of performing multiple lockin measurements simultaneously and for such measurements, clock synchronization is very important.  Figure~\ref{sync} compares two measurements, where two sine waveforms were added in a summing amplifier, sampled and Fourier transformed.  We see that the removal of the synchronization between the synthesis clock and the sampling clock causes the appearance of Fourier leakage.  This leakage dramatically raises the measurement floor in the vicinity of the drive frequencies, below which real signals with good signal-to-noise ratio could be hiding.   Synchronization establishes one master clock from which a common reference signal is derived, and is thus a key aspect of digital lockin measurements.

The lockin method is often used to probe the response of a two-port system to a pure drive tone.  The drive applied at one port, $V_D\cos(\omega_D t)$, is used as a reference for the measurement of two channels of response, $V(t)$, at the other port.  The lockin measurement is then realized by multiplying two unity-amplitude copies of the drive signal, one being phase-shifted by $\pi/2$, and integrating each for a time $T$.  The result will be the Fourier cosine and sine components of the response in a frequency bandwidth $1/T$, centered at the reference frequency $\omega_D$,

\begin{eqnarray}
V_{x}= \frac{1}{T} \int_0^T V(t) \cos(\omega_D t)dt\,,
\\
V_{y}= \frac{1}{T} \int_0^T V(t) \sin(\omega_D t)dt\,.
\end{eqnarray}

In the digital world we can realize this measurement by sampling the response at discrete times and calculating a Fourier sum. The cosine and sine quadratures $V_x$ and $V_y$ can be alternatively represented as amplitude $\vert V \vert = \sqrt{V_x^2 + V_y^2}$ and phase $\theta$, where $\tan \theta = {V_y}/{V_x}$.  The reference signal is taken to have zero phase, and the response phase $\theta$ is the phase with respect to this zero. 

This approach to measuring signals is natural for linear systems driven by a pure tone, where the response will only be at the drive frequency.   For nonlinear systems the response to one pure drive tone will include harmonics, or response at integer multiples of the drive frequency. The lockin amplifier can be easily extended to include harmonic analysis by simply forming two new measurement channels for each harmonic frequency desired.  In this manner, the phase of the harmonics with respect to the reference signal is unambiguously defined.

When a nonlinear system is driven with two pure tones at $\omega_1$ and $\omega_2$, the response contains not only harmonics of the drive tones, but also intermodulation products (mixing products) of the drive tones, occurring at frequencies which are integer linear combinations of the two drive frequencies,
$\omega_{\rm IMP}= k_1 \omega_1 + k_2 \omega_2$ where $k_1, k_2  \in \mathbb{Z}  $.  For the intermodulation response, which does not occur at a drive frequency or a harmonic, there is no reference signal and an unambiguous definition of phase cannot be found.  We can however measure the phase of intermodulation products by restricting our measurement so that all drive tones occur at frequencies which are integer multiples of some base frequency, which we call $\Delta \omega$.  In this case all intermodulation products occur at integer multiples of $\Delta \omega$, and $\Delta \omega$ can be used as the reference frequency in the lockin calculation.

\section*{Intermodulation Lockin Analyzer}

\begin{figure}
\includegraphics{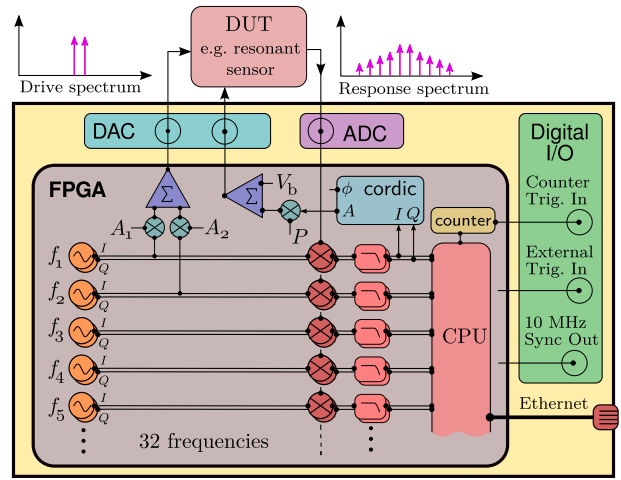}
\caption{A schematic showing 
the functionality of the digital Intermodulation Lockin Analyzer (ImLA). The frequencies ($f_i$), drive amplitudes ($A_1$, $A_2$), feedback gain ($P$) and bias ($V_\mathrm{b}$) can all be controlled via the ethernet interface. For each of the 32 frequencies both the inphase ($I$) and the quadrature ($Q$) component is generated and used in a lockin calculation.}
\label{schematic}
\end{figure}

We used a field-programmable gate array with one channel of A/D input (12 bit) and two channels of D/A output (16 bit) to implement an intermodulation lockin measurement, where all relevant frequencies are integer multiples of $\Delta \omega$.  Schematically depicted in fig.~\ref{schematic}, the Intermodulation Lockin Analyzer (ImLA) synthesizes two user-defined drive tones and calculates both quadratures of the intermodulation response at 32 user-defined intermodulation frequencies.  The user communicates with the ImLA via an ethernet connection so the instrument can easily be integrated into measurement systems using a wide variety of software platforms (e.g. LabView\textregistered, MATLAB\textregistered, Python\textregistered, etc.).  

In its present form the ImLA samples at 61.4 MSa/s.  64 Fourier sums are calculated in parallel by accumulating every sample during a period of time $T$, which is the inverse of the user-defined measurement bandwidth $1/T$.  The quadrature response at 32 user-defined frequencies is thus acquired and sent over ethernet to a computer.  For one of the 32 frequencies the amplitude and phase are calculated by the CORDIC algorithm \cite{volder:cordic:59} in the FPGA, so that these signals can be used for real-time feedback.  The update rate of the feedback can be selected up to $1024/T$.  This feature is useful in AFM systems where fast feedback is needed to control the scanning probe's height above the surface.  

The Intermodulation lockin can also be run in a time-domain mode, where the response is continuously downsampled to 3.9 MSa/s and streamed over ethernet without gaps in the data.   This mode is useful for detailed study of the entire response spectrum.  For synchronization with other measurement instruments, the ImLA has a 10 MHz clock OUT signal and an external trigger input which will reset the calculation of the Fourier sums, effectively defining the start of measurement.   Many of the ImLAs specific features and performance specifications can be modified by reconfiguration of the firmware on the FPGA.

The ImLA's ability to measure nonlinearity in a Device Under Test (DUT) is limited by its own internal nonlinearity.   We tested this limit by sending a maximum amplitude drive signal directly from the OUT port to the IN port, to measure an internal Total Harmonic Distortion (THD) and IMD, which were both less than $-75$~dB.  Under these test conditions we also find that the dynamic reserve at $1$~kHz measurement bandwidth is $80$~dB. 

\section*{Intermodulation Atomic Force Microscopy}

\begin{figure}
\includegraphics[width=8cm]{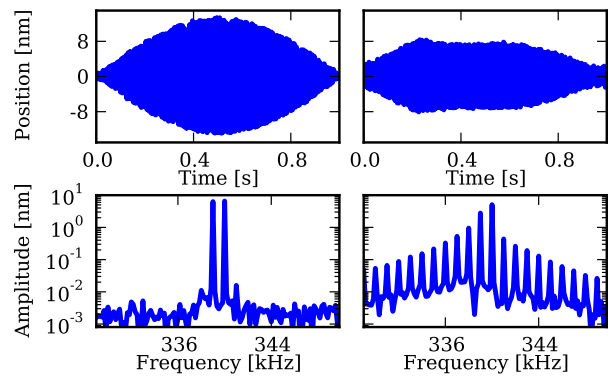}
\caption{ Time domain (upper panels) and frequency domain (lower panels) of free cantilever (left panels) and an engaged cantilever (right panels) in Intermodulation AFM (enhanced online\cite{IntermodLockinSupplement}).  The time domain shows the sampled response over only one beat of the drive wavelet.  The frequency domain shows the FFT of the sampled data over 10 consecutive drive beats.  The small measurement bandwidth used in the FFT allows one to observe how the noise level in between the intermodulation peaks changes while approaching the surface.  }
\label{spectra}
\end{figure}

Figure~\ref{spectra} demonstrates the use of the ImLA in Intermodulation Atomic Force Microscopy \cite{Platz:ImAFM:08} where IMD in a mechanical oscillator is measured.  Typically the first flexural eigenmode of the cantilever is excited, which often has a sharp resonance around 350 kHz with quality factor $Q \sim$ 600 in air.  The ImLA is used to drive the cantilever with two tones separated by $ \sim 500$ Hz, centered around resonance.  The cantilever response is detected by optical means in the AFM, and the output of a photo-diode circuit is analyzed by the ImLA.  

The free cantilever oscillating well above the surface is a linear system and the response to two drive tones closely spaced in frequency will be a beating waveform in the time domain (fig.~\ref{spectra} shows one beat).  When the oscillating cantilever approaches an SiO$_2$ surface, the sharp tip on the end of the cantilever experiences a nonlinear tip-surface force, which creates intermodulation of the drive tones.  Figure~\ref{spectra} shows two frames from a movie of an approach-retract measurement, which can be seen online\cite{IntermodLockinSupplement}.  In these measurements the ImLA was used in time-domain mode and the spectra were calculated by Fourier transform after the measurement.  

When the AFM is scanning over a surface, speed in analysis is required, and the ImLA is used in lockin mode, where the intermodulation response is captured only at the peaks in fig.~\ref{spectra} (lower right panel).  While scanning, multiple images can be viewed by plotting the amplitude or phase at the various intermodulation frequencies as a function of tip position \cite{Platz:PhaseImage:2010}.  From the intermodulation spectrum measured at each pixel of the image, one can reconstruct the tip-surface interaction force \cite{Hutter:Reconstructing:2010}, and reveal the nano-scale mechanical properties of the surface.  This nonlinear interaction force is the information which can be measured with dynamic AFM, and one can view the IMD measurement not only as a sensitive probe of this nonlinear interaction, but also as an efficient means of compressing information in the sampled data.  By locking in on the intermodulation frequencies we extract for storage only the essential amplitudes and phases in the response, disregarding only noise. 

\section*{Multi-Frequency Intermodulation Analysis}

Thus far we have considered a drive consisting of two pure tones and a measurement of the intermodulation products of these two tones.  A natural extension of this technique would be to include arbitrary drive waveforms.  For example, we could synthesize a drive wavelet other than a simple beating waveform, such as a chirplet formed by combining several frequencies in a band \cite{Jesse:BandExcitation:07}.  The drive waveform should be built from a superposition of pure tones where the frequencies are taken from a set of frequencies formed by the integer multiples of $\Delta \omega$.  A lockin method will thus be possible since intermodulation response will occur only at integer multiples of $\Delta \omega$.  In this case, the analysis of the response is identical to that described here, and the architecture of the ImLA is well suited for such an extension.  However, for multiple drive tones, the measured intermodulation would include all possible mixing products of the many drive frequencies.  Nevertheless, the theoretical approach based on polynomial representation of the nonlinearity to create a linear mapping of the intermodulation products in frequency space \cite{Hutter:Reconstructing:2010} should be able to handle this complication.  It is possible that suitably chosen drive wavelets could be advantageous for the intermodulation spectral technique, depending on the application and the information one wishes to extract from the intermodulation measurement.    

In summary, we described a digital incarnation of a lockin measurement instrument specifically designed to measure the intermodulation of pure tones by nonlinearity.  This Intermodulation Lockin Analyzer (ImLA) can find use in many measurement applications where one wishes to make a detailed model of the nonlinearity of a device under test.  The intermodulation spectral technique is particularly useful in applications where the perturbation of a high quality factor resonance is employed for sensing.   We demonstrated one such an application in the field of dynamic Atomic Force Microscopy, however numerous sensors based on this measurement principle could use the intermodulation lockin to gather more information than is otherwise possible with standard lockin techniques based on a single drive frequency.   

\begin{acknowledgments}
This work was supported by The Swedish Research Council (VR) and The Swedish Governmental Agency for Innovation Systems (VINNOVA).  
\end{acknowledgments}

\bibliography{David}

\end{document}